\documentclass[%
 reprint,
 amsmath,amssymb,
 aps, showkeys,
prl,
]{revtex4-2}

\usepackage{graphicx}
\usepackage{dcolumn}
\usepackage{bm}
\usepackage{hyperref}
\usepackage{lipsum}
\usepackage{subfigure}
\usepackage{amsmath}

\graphicspath{ {./images/} }

\hypersetup{
    colorlinks = true,
    linkcolor = blue,
    citecolor = blue,
    urlcolor = blue,
}

\begin{document}

\preprint{APS/123-QED}

\title{On the feasibility of Ohmically heated negative triangularity tokamak power plants}

\author{A. Balestri$^{1}$}
\author{J. Ball$^1$}
\author{S. Coda$^1$}

\affiliation{$^1$École Polytechnique Fédérale de Lausanne (EPFL), Swiss Plasma Center (SPC), 1015 Lausanne, Switzerland}


\begin{abstract}

Negative triangularity tokamak plasmas feature naturally enhanced confinement in the so-called L-mode regime, irrespective of the power of external heating. This is in contrast to conventional scenarios, which require exceeding a given heating power threshold to induce a discrete transition to a regime of enhanced confinement called H-mode. H-mode is, however, subject to problematic instabilities and additionally suffers from confinement degradation with increasing external heating. Using simple zero dimensional power balance and standard empirical scaling laws for confinement, we analyze the impact of external heating on several different reactor-relevant devices (i.e. SPARC, MANTA, ITER and DEMO). We compare the nominal externally heated scenarios with corresponding negative triangularity cases without external heating. For devices with sufficiently high magnetic field and/or fusion gain, the internally (Ohmically) heated negative triangularity versions achieve better performance. We conclude that Ohmically heating a negative triangularity power plant is an attractive option meriting further investigation.

\end{abstract}

\keywords{fusion plasma, negative triangularity, Ohmic heating, fusion power plant, ignition}
\maketitle

\paragraph*{Introduction} The vast majority of designs for reactor-relevant tokamaks employ Positive Triangularity (PT) plasma shaping and powerful heating systems to generate large quantities of fusion power. ``Positive triangularity'' refers to a D-shaped poloidal cross-section in which the nose points to the low-field side of the torus. Powerful heating systems are required as these designs typically rely on the plasma entering the H-mode (High confinement mode) regime. Crucially, this requires one to exceed a heating power threshold, which can be tens of megawatts \cite{Martin_2008,Behn_2015}. This drives the need for complex and expensive heating systems that require a large amount of electricity. Additionally, while H-mode confinement is better than its L-mode (Low confinement mode) counterpart, confinement decreases with increasing heating. For these reasons, reactor designs typically operate with the minimal amount of heating needed to trigger H-mode \cite{SPARC}.

While H-mode achieves excellent energy confinement, it has a serious drawback. The edge of H-mode plasmas is not stable, as it suffers from periodic virulent instabilities called Edge Localized Modes (ELMs). In future power plants, ELMs are expected to release enormous fluxes of energy to the plasma facing components of a reactor, exceeding material survivability limits. This has deepened interest in an attractive alternative: Negative Triangularity (NT). The poloidal cross-section of this plasma is shaped like a ``\reflectbox{D}'', instead of ``D''. In the last two decades, experiments \cite{8,9,Coda_2022,Balestri_2024_DTT,7,Paz-Soldan_2024,ASDEX,Aucone_2024} and simulations \cite{Marinoni_2009,5,Merlo_Jenko_2023,Balestri_2024,digiannatale2024,Ball_2023} have revealed several advantages of NT. First, NT plasmas are {\it not} able to enter H-mode \cite{nelson2024_2}, even at high heating power. Instead they remain in L-mode, the most robust ELM-free scenario. Second, NT still features a reduction of turbulent transport, which allows a NT plasma to sustain strong plasma gradients in the outer core and achieve central core pressures similar to a standard PT H-mode. Thus, NT achieves H-mode-like global confinement, while remaining in L-mode and avoiding ELMs. Additionally, this means that the constraint of exceeding the L-H power threshold vanishes and external heating systems are no longer required to improve confinement \cite{Nelson_PRL2023}. In this paper, we will consider the question ``What is the ideal amount of heating power for a NT tokamak?'' In particular, we will consider NT plasmas with the minimum amount of heating possible: only the Ohmic heating that naturally arises from the toroidal plasma current. 

\paragraph*{Methodology}

To assess the performance of various scenarios, we use the simple zero-dimensional (0D) power balance equation
    \begin{equation}
    \langle \frac{dW}{dt}\rangle=\langle p_{\alpha}\rangle+\langle p_{\Omega}\rangle+\langle p_{ext}\rangle-\langle p_{rad}\rangle-\langle p_{loss}\rangle,
    \label{0D_general}
\end{equation}
where $dW/dt$ is the time variation of thermal energy density $W$ stored in the plasma, $p_\alpha$ is the power density released by the $\alpha$ particles produced by fusion reactions, $p_{\Omega}$ is the Ohmic heating power density generated by the electric current flowing in the plasma, $p_{ext}$ is the heating power density injected by external systems (e.g. neutral beam injection, radiofrequency heating), $p_{rad}$ is the power density radiated by the plasma and $p_{loss}=W/\tau_E$ is the power density lost through transport (primarily turbulence-driven transport). To avoid confusion, power \textit{density} is expressed with lower case $p$, while power is $P$.  The definitions and formulas for each power density can be found in appendix A.

The $\langle \ldots \rangle$ operator is a volume average over the entire plasma, which allows us to consider profile effects. This requires knowledge of how temperature $T$ and density $n$ vary with minor radius $r$, which is set by transport processes that are beyond the simple analysis in this work. Hence, we will assume simple analytical profiles which allow us to write volume averaged quantities as  $\langle f\rangle=\frac{\int dV f(r)}{\int dV}=\frac{f_0}{1+\alpha_f}$, where the integral is performed over the plasma volume $V$. Here $f$ is a placeholder for $T$ or $n$, $f_0$ is its value at $r=0$ and $\alpha_f$ is what we refer to as a \textit{peaking factor}. For $\alpha_T$ and $\alpha_n$ we will use typical values from theory and experiment. To keep the analysis as simple as possible, we assume that electrons and ions have equal densities and temperatures, so that the stored energy can be expressed as $W=3nT$. We restrict our analysis to plasmas composed of an equal mixture of deuterium and tritium.

The equilibrium state is set by $dW/dt=0$, i.e. when the sources are balanced by the sinks in the power balance equation. We will consider density to be a free parameter, as it can be controlled by changing the fueling. Thus, the condition $dW/dt=0$ sets the temperature(s) that can be achieved by the plasma in steady-state. In this work, we will determine the steady-state temperatures and compute the associated fusion power $P_{fus}=5P_\alpha$ and fusion gain $Q=P_{fus}/(P_\Omega+P_{ext})$.

The only quantity that we have not yet specified in equation \eqref{0D_general} is the energy confinement time $\tau_E$, for which we use the most widely accepted scaling laws with one caveat. As Ohmic heating is key in this work, we cannot simply use scaling laws based primarily on externally heated plasmas. As showed in many publications \cite{Rice_2020}, confinement in Ohmic plasmas scales linearly with density in the so-called Linear Ohmic Confinement regime and transitions to Saturated Ohmic Confinement (a scaling law approaching that for externally heated plasmas) once a critical density is reached (i.e. the LOC-SOC transition). However, we also cannot exclusively use Ohmic scaling laws because the $\alpha$ heating becomes significant at high performance. This represents ``external'' heating as it is a distinct source of heating absent from all past experimental Ohmic discharges. For these reasons, we unify Ohmic and externally heated energy confinement scaling laws following \cite{lampis1986basic} to write
\begin{equation}
    \tau_E=\tau_{\Omega}\left(\frac{p_{\Omega}}{p}\right)+\tau_{heat}\left(1-\frac{p_{\Omega}}{p}\right),
    \label{tauCombined}
\end{equation}
where $\tau_\Omega$ is the confinement time given by the Linear Ohmic Confinement scaling, $\tau_{heat}$ is the scaling for the heated part and $p$ is the total injected power density (i.e. $p=p_\alpha+p_\Omega+p_{ext}$). This formula allows a natural transition from a purely Ohmic to a heated scenario. In the Ohmic part of equation \eqref{tauCombined}, we model the transition from the LOC to the SOC regime by using the experimental critical density ($n_{crit}=0.65M^{0.5}B/(q_{95}R)10^{20}m^{-3}$) \cite{Rice_2020} above which $\tau_\Omega$ is clamped to a fixed value given by the LOC-SOC critical density.  As detailed in appendix B, we use the standard ITER and neo-Alcator scaling laws for $\tau_{heat}$ and $\tau_\Omega$ respectively. 

\paragraph*{Analytic results} To better understand the exact numerical solutions to equation \eqref{0D_general} presented in the next section, we will first perform a simplified analytic calculation. We seek the optimal amount of external heating $P_{ext}$ to maximize the fusion gain $Q = P_{fus}/P_{ext}$. To find this, we must determine the conditions for which
\begin{equation}
    \frac{d Q}{d P_{ext}} = 0 \Rightarrow \frac{d P_{fus}}{d P_{ext}} = Q \Rightarrow  \frac{d P_{fus}}{d T_0} \frac{d T_0}{d P_{ext}} = Q,
    \label{A}
\end{equation}
is satisfied. The dependence of the on-axis temperature $T_0$ on the external heating power $P_{ext}$ is complicated as it requires solving equilibrium power balance (i.e. $dW/dt = 0$). To enable a simple solution, we take the strongly heated limit of equation \eqref{0D_general} (i.e. neglecting the radiated and Ohmic powers) to find
\begin{equation}
     0 = \frac{P_{fus}}{5} + P_{ext} - \frac{3 \langle n T\rangle V}{\tau_{heat}},
    \label{B}
\end{equation}
where $\tau_{heat}$ is given by either equation \eqref{tau98} or \eqref{tau89}. Though we neglected the Ohmic heating, this calculation will still be useful as it will tell us when strong external heating is beneficial and when it is best to turn the external heating systems down as low as possible. Taking the derivative of equation \eqref{B} with respect to $P_{ext}$, substituting back in equation \eqref{B} and using the definition of Q allows us to calculate (after a lengthy derivation) the optimal amount of heating power to be
\begin{equation}
    P_{ext} = \frac{1 - d}{Q} T_0 \frac{d P_{fus}}{d T_0},
    \label{C}
\end{equation}
where $d$ is the heating power degradation exponent from $\tau_{heat}\propto P_{ext}^{-d}$. This can also be recast as an equation for the ideal operating temperature
\begin{equation}
    \frac{T_0}{P_{fus}}\frac{d P_{fus}}{d T_0} = \frac{1}{1 - d},
    \label{D}
\end{equation}
which is interesting as it is independent of the parameters of a given device. The left side depends on constants of nature and the peaking factors, while the right side depends only on the strength of power degradation $d$. Letting $\alpha_n=\alpha_T=0$, equation \eqref{D} tells us that to maximize $Q$ all devices operating in H-mode ($d=0.69$) should target an on-axis temperature of $T_0 = 6.4\, keV$, while all devices operating in L-mode ($d=0.5$) should target $T_0 = 13.6\, keV$. Plugging these temperatures into equation \eqref{C} allows us to determine the heating power corresponding to this optimum. We see that this optimal heating power does depend on the machine through the power gain $Q$ that the machine would achieve at that temperature. Equation \eqref{C} tells us that low performance devices (i.e. low $Q$) benefit from a lot of external heating power, as it enables them to reach the optimal temperature that maximizes $Q$. However, powerful heating systems are less useful in high performance devices like a power plant, since such machines are capable enough to operate at high $Q$. In the limit of devices that can ignite, there is clearly no need for any external heating systems.

The above calculation is insightful for interpreting trends in the numerical results in the next section. However, the assumptions we made (in particular neglecting the radiated power) do limit its numerical accuracy. While the quantitative values for the optimal temperatures from equation \eqref{D} appear to hold fairly well, the solutions for the ideal heating power from equation \eqref{C} can be substantially off.

\paragraph*{Numerical results} Using equation \eqref{0D_general}, we can estimate the operating point of various tokamak designs to understand the impact of external heating on NT plasmas, as well as compare NT with existing PT H-mode designs. We will consider the four reactor-relevant tokamaks shown in table \ref{param}: MANTA \cite{MANTA_2024,Miller_2024,Wilson_2024}, SPARC \cite{SPARC}, ITER \cite{ITER2007} and the European DEMO \cite{DEMO2019}.

\begin{table}[h]
\begin{tabular}{l|cccc}
\toprule
 & MANTA & SPARC & ITER & DEMO \\\hline
$I_p [MA]$ &   10    &   8.7    &   15 & 19.6   \\
$B_T [T]$ &    11   &   12.2    &   5.3  & 5.7 \\
$a [m]$ &    1.2   &   0.57    &   2.0  & 3.0 \\
$R [m]$ &   4.55    &    1.85   &   6.2  & 9.1 \\
$\kappa_{95}$ &   1.4    &    1.7   &   1.6  & 1.6 \\
$\delta_{95}$ &   -0.5    &   0.45    &   0.33  & 0.33 \\
$Z_{eff}$ &   1.5    &    1.5   &   1.5  & 1.5 \\
$\alpha_T$ &   1.5    &   1.5    &    0.6  & 1.5\\
$\alpha_n$ &   0.3    &   0.33    &    0.2 & 0.3\\
$P_{ext}$ &   40    &   11    &    40  & 50\\
$Q^{target}$ &   15    &   11    &    10 & 40 \\\toprule 
\end{tabular}
\caption{Main parameters for the considered tokamaks.}
\label{param}
\end{table}

These machines were chosen because they represent different parts of parameter space. SPARC is relatively small with a very large magnetic field, MANTA is medium-sized with a large magnetic field, ITER is large with relatively low magnetic field and DEMO is very large with low magnetic field. MANTA is designed for NT, while all the others use PT. Note that we use standard H-mode density peaking  factors for all cases, which will tend to underpredict the performance of the broader NT profiles.

\begin{figure*}
    \centering
    \begin{subfigure}[]
    {\includegraphics[width=0.32\textwidth]{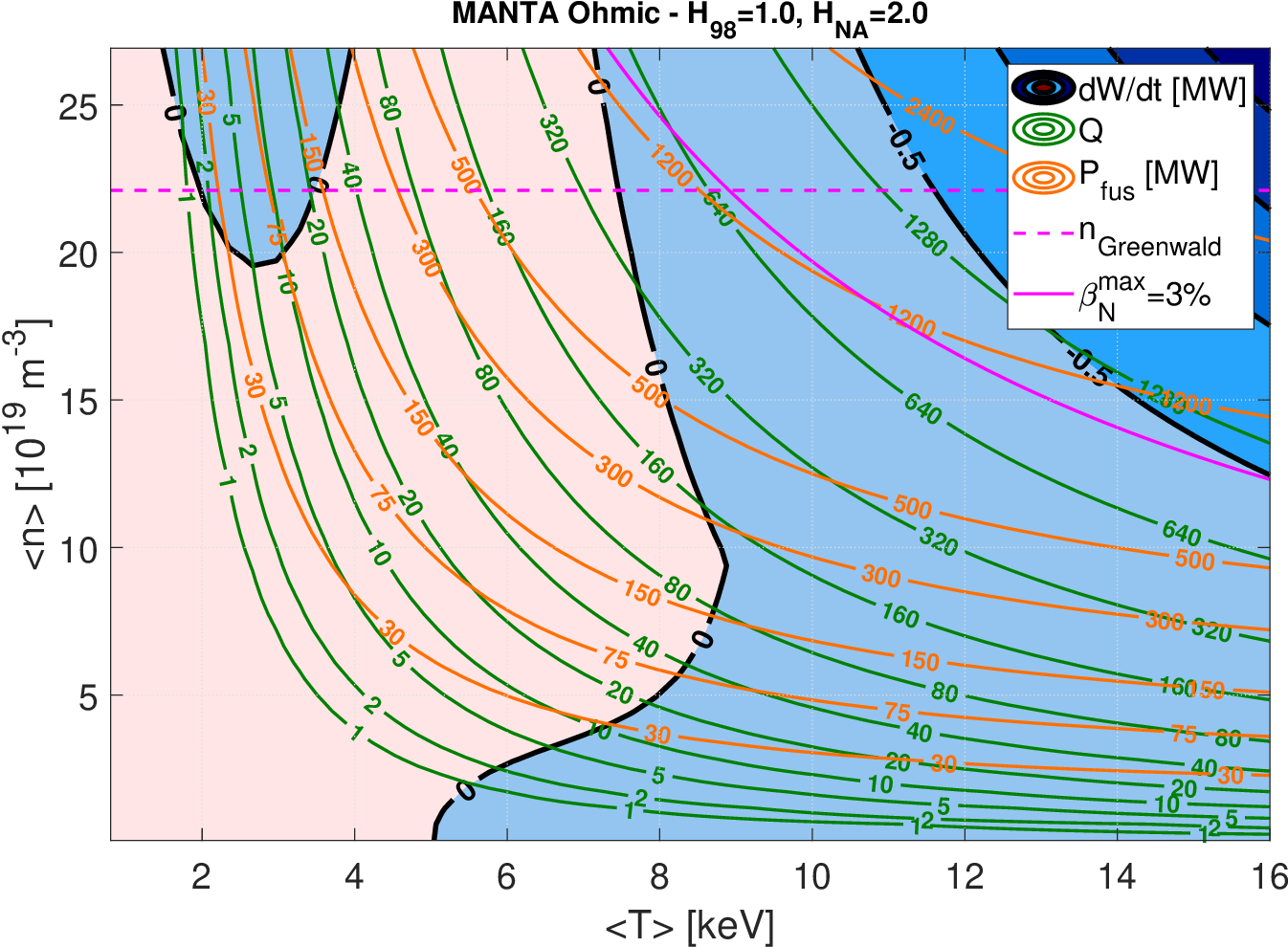}}
    \end{subfigure}
    \begin{subfigure}[]
    {\includegraphics[width=0.32\textwidth]{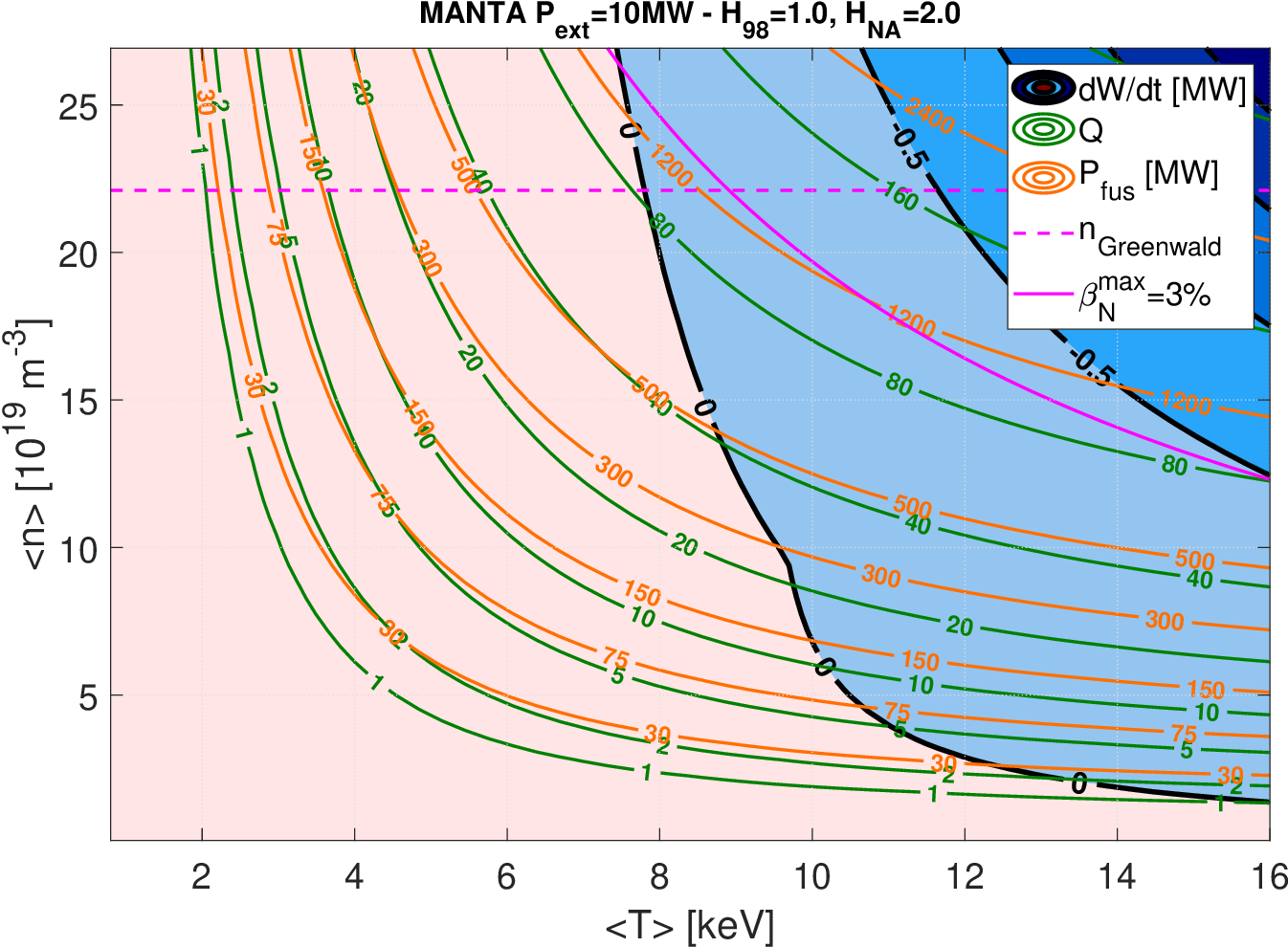}}
    \end{subfigure}
    \begin{subfigure}[]
    {\includegraphics[width=0.32\textwidth]{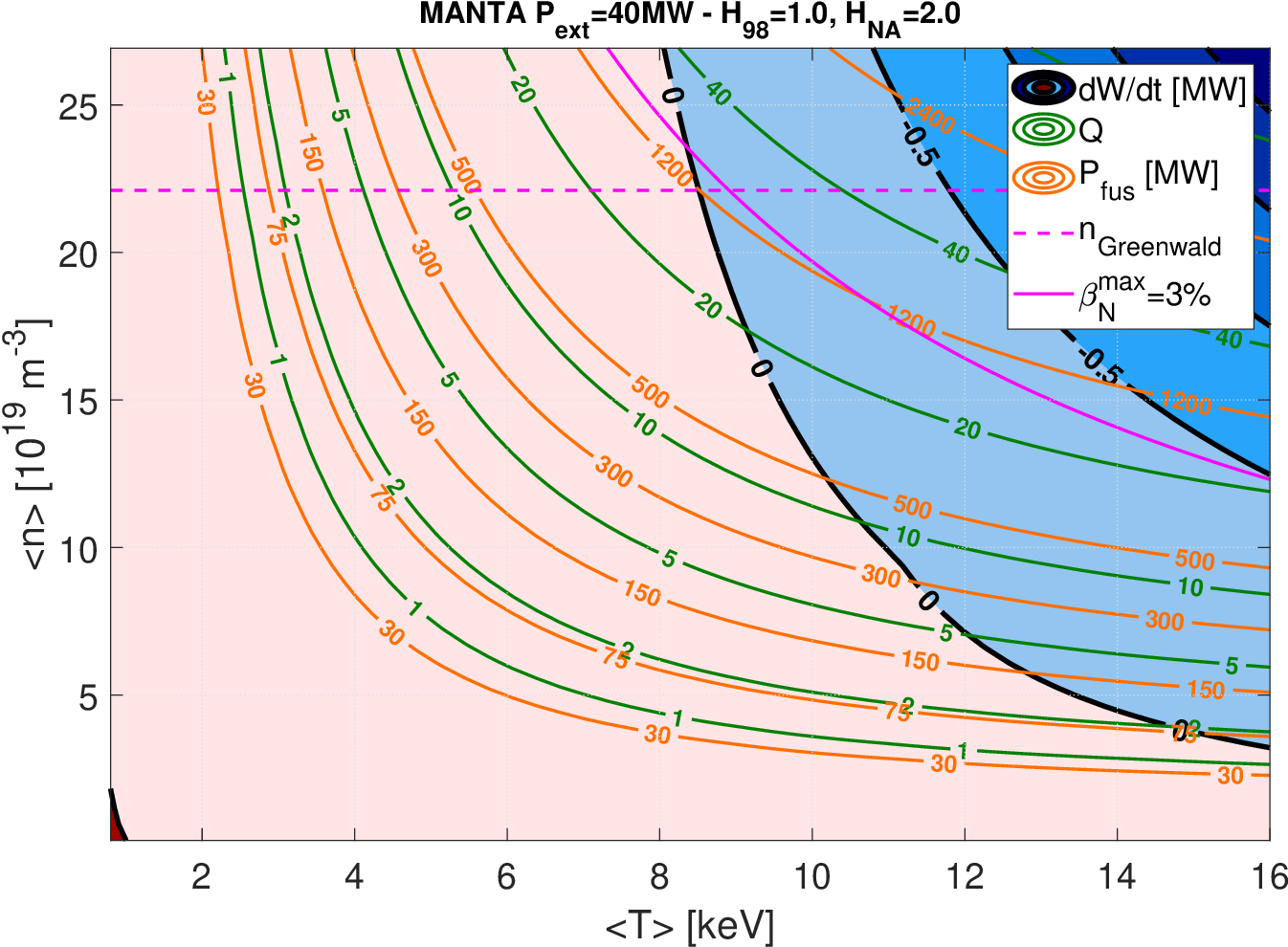}}
    \end{subfigure}
    \begin{subfigure}[]
    {\includegraphics[width=0.32\textwidth]{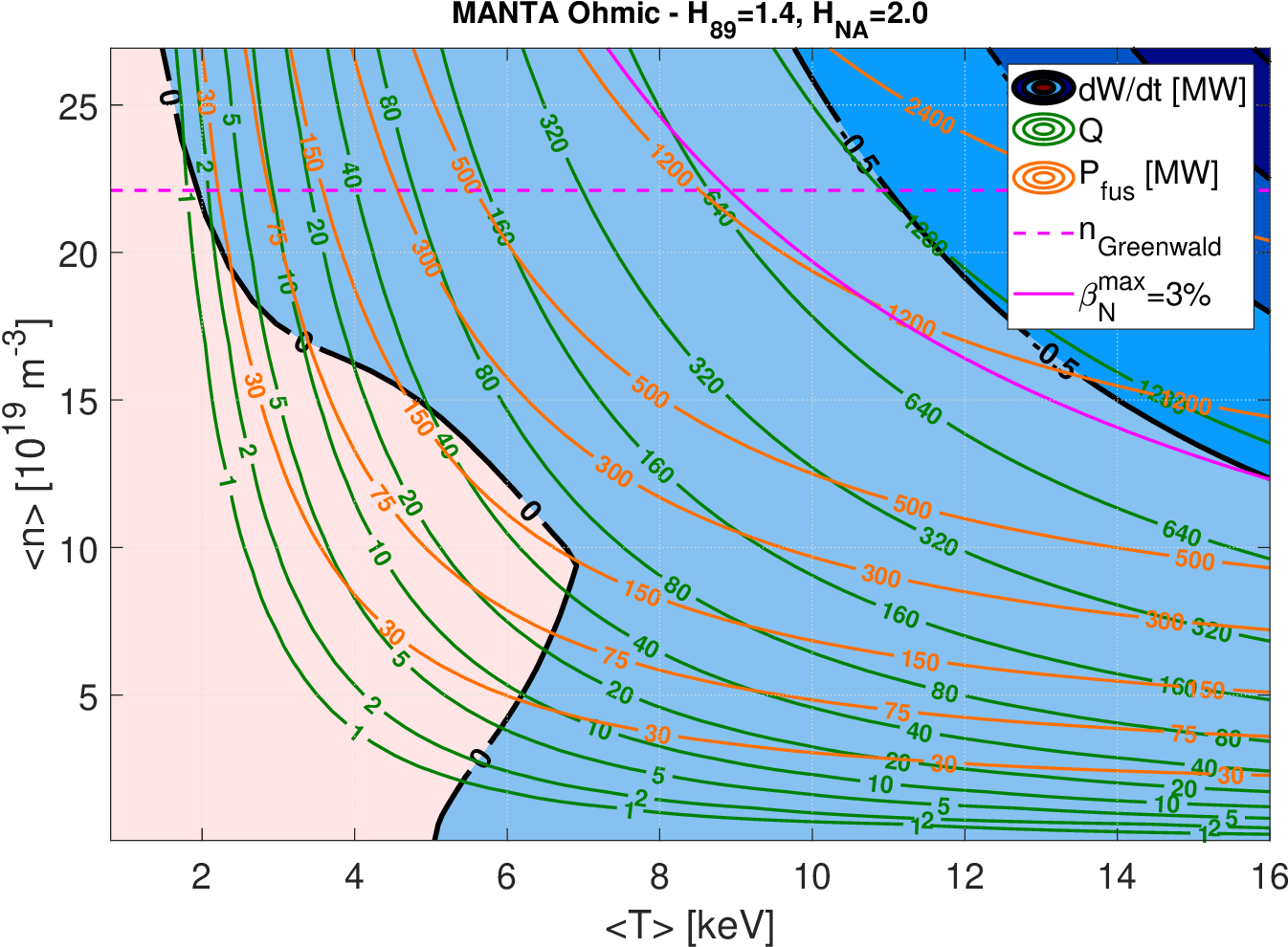}}
    \end{subfigure}
    \begin{subfigure}[]
    {\includegraphics[width=0.32\textwidth]{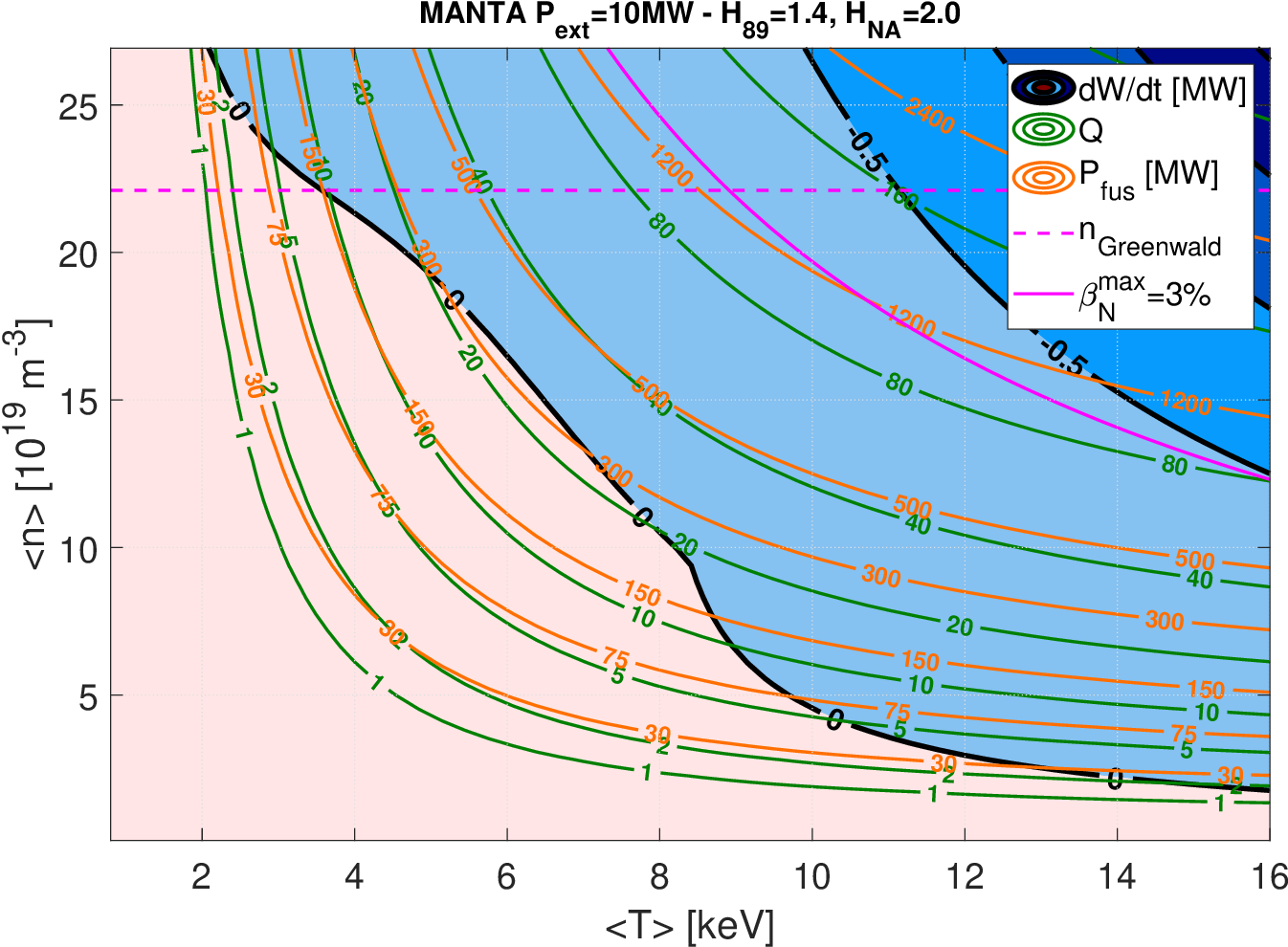}}
    \end{subfigure}
    \begin{subfigure}[]
    {\includegraphics[width=0.32\textwidth]{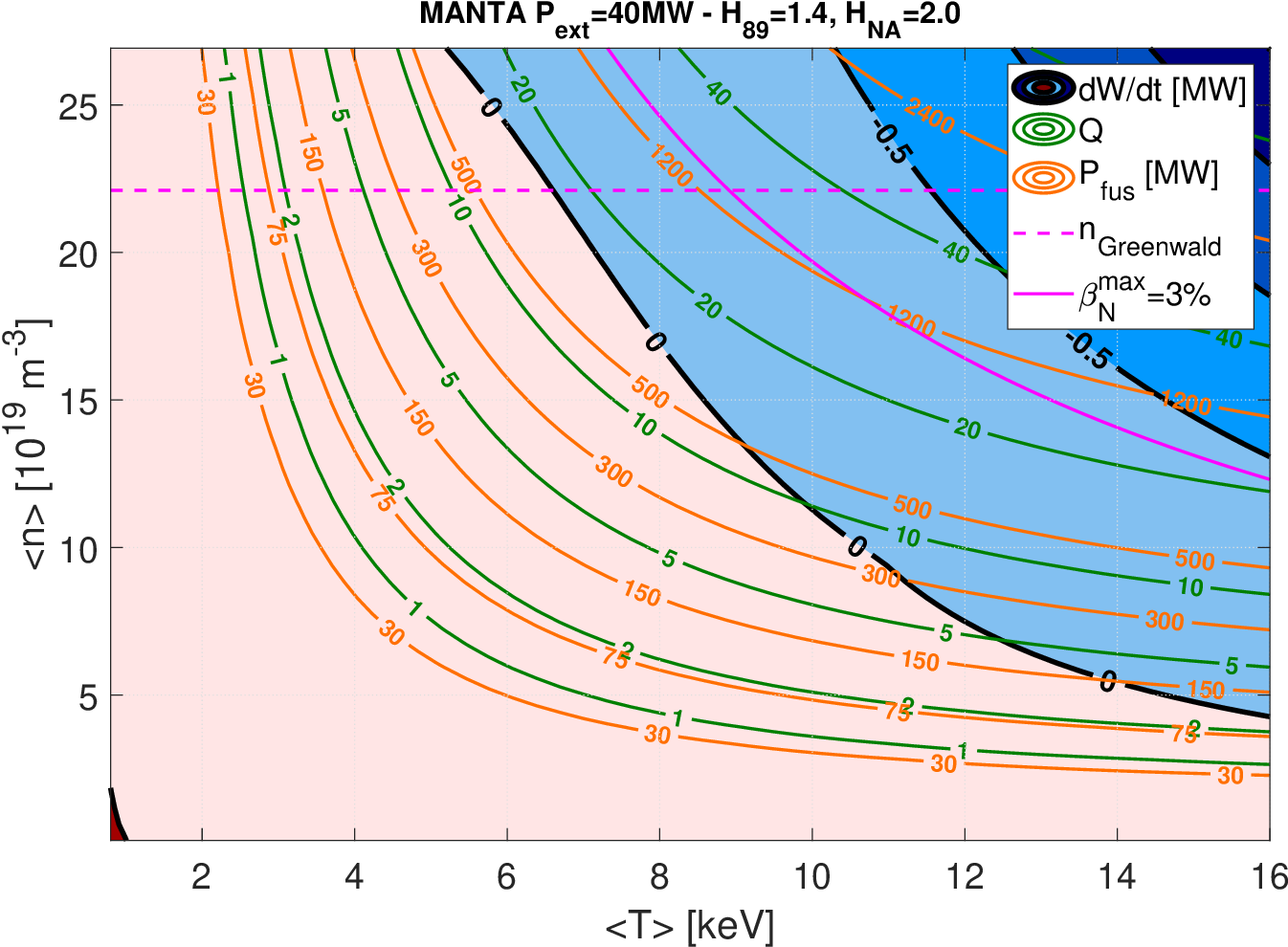}}
    \end{subfigure}
    \caption{Contour plots of $dW/dt$ in temperature-density space for MANTA scenarios with different values of external heating power and confinement time scalings. The green contours indicate constant fusion gain $Q$, the orange contours indicate constant fusion power $P_{fus}$, the dashed magenta line indicates the Greenwald density limit and the solid magenta curve indicates the Troyon $\beta_N$ limit.}
    \label{MANTA}
\end{figure*}

\begin{figure*}
    \centering
    \begin{subfigure}[]
    {\includegraphics[width=0.23\textwidth]{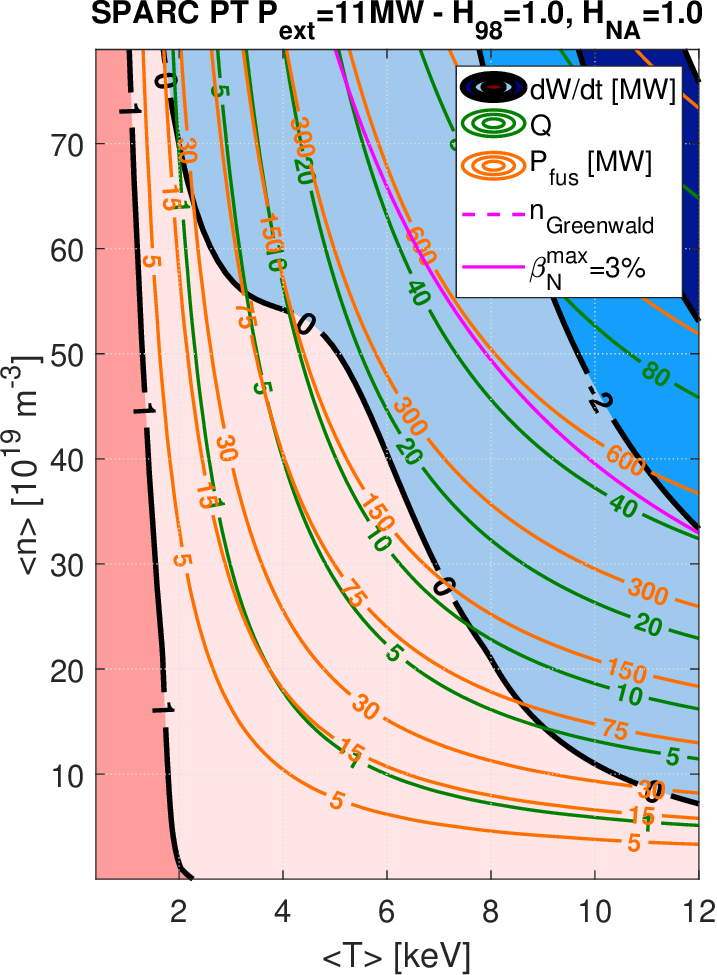}}
    \end{subfigure}
    \begin{subfigure}[]
    {\includegraphics[width=0.23\textwidth]{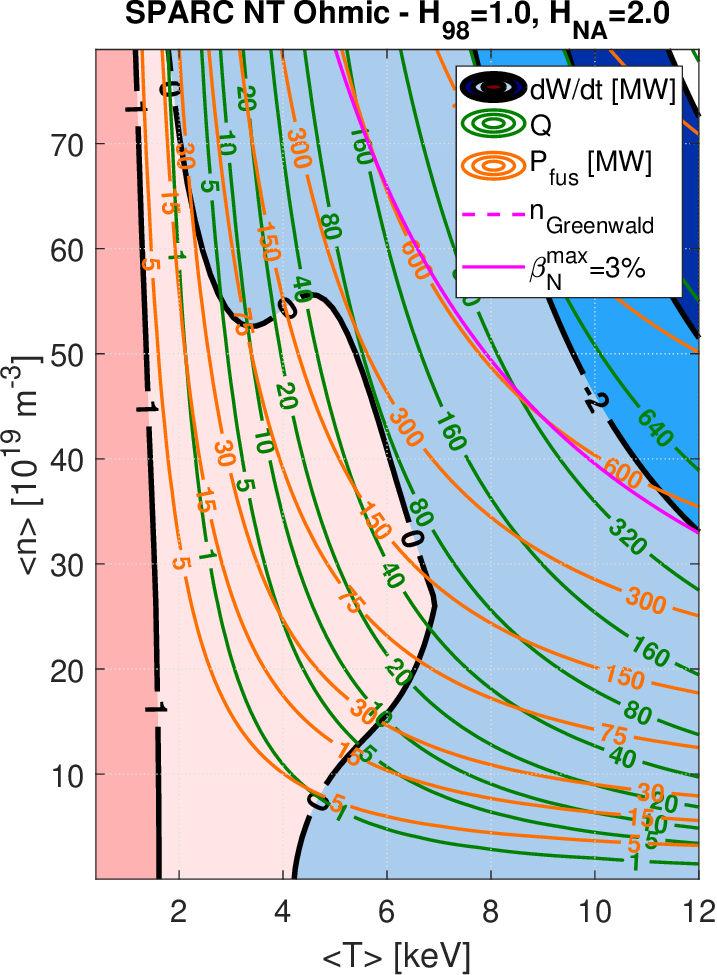}}
    \end{subfigure}
    \begin{subfigure}[]
    {\includegraphics[width=0.23\textwidth]{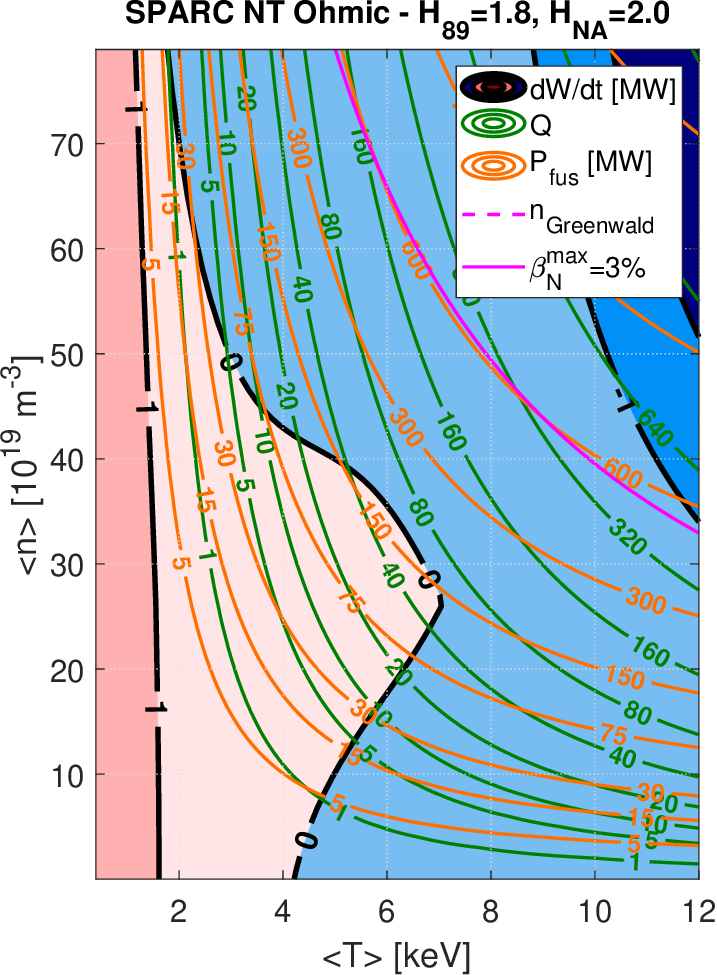}}
    \end{subfigure}
    \begin{subfigure}[]
    {\includegraphics[width=0.23\textwidth]{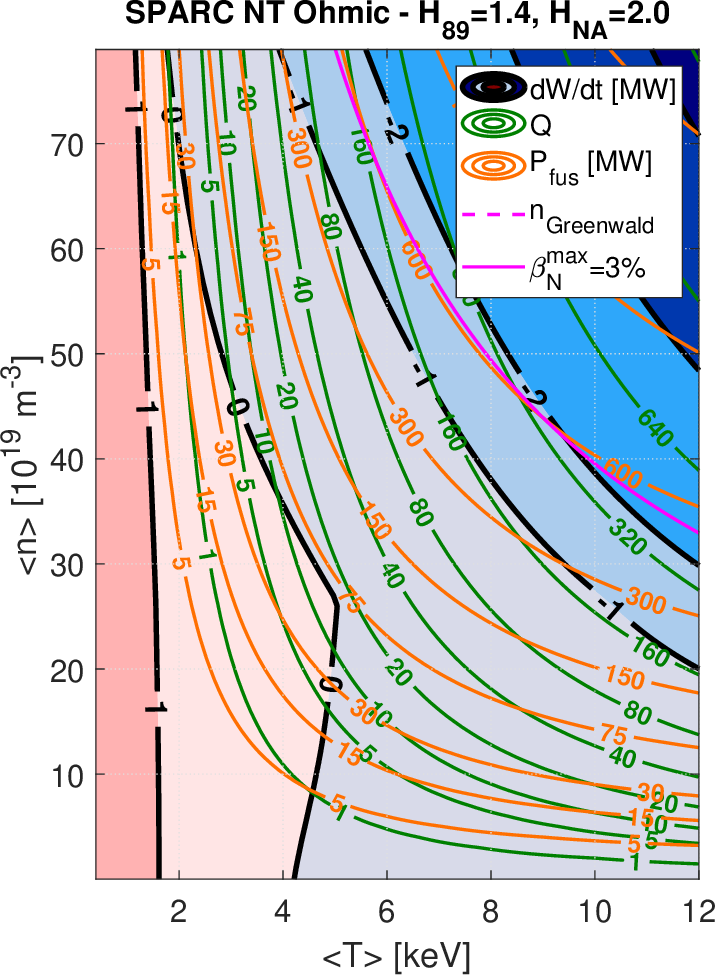}}
    \end{subfigure}
    \caption{Contour plots of $dW/dt$ in temperature-density space for different SPARC PT (a) and NT (b,c,d) scenarios with H-mode or enhanced L-mode confinement time scalings. The various contour curves have the same meaning as in figure \ref{MANTA}.}
    \label{SPARC}
\end{figure*}

\begin{figure*}
    \centering
    \begin{subfigure}[]
    {\includegraphics[width=0.24\textwidth]{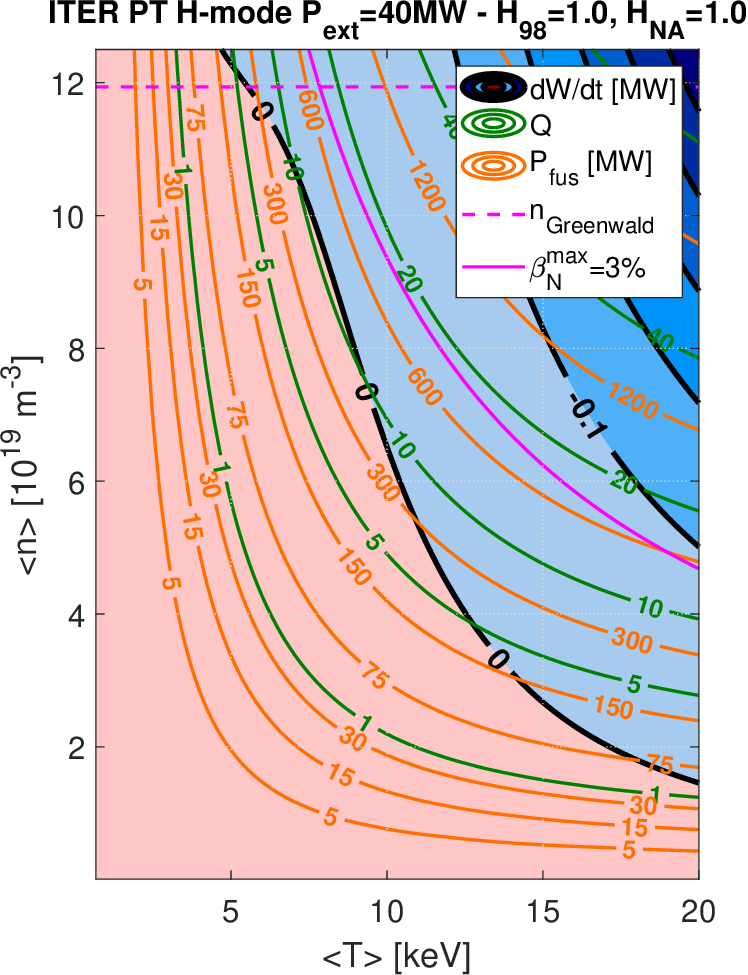}}
    \end{subfigure}
    \begin{subfigure}[]
    {\includegraphics[width=0.23\textwidth]{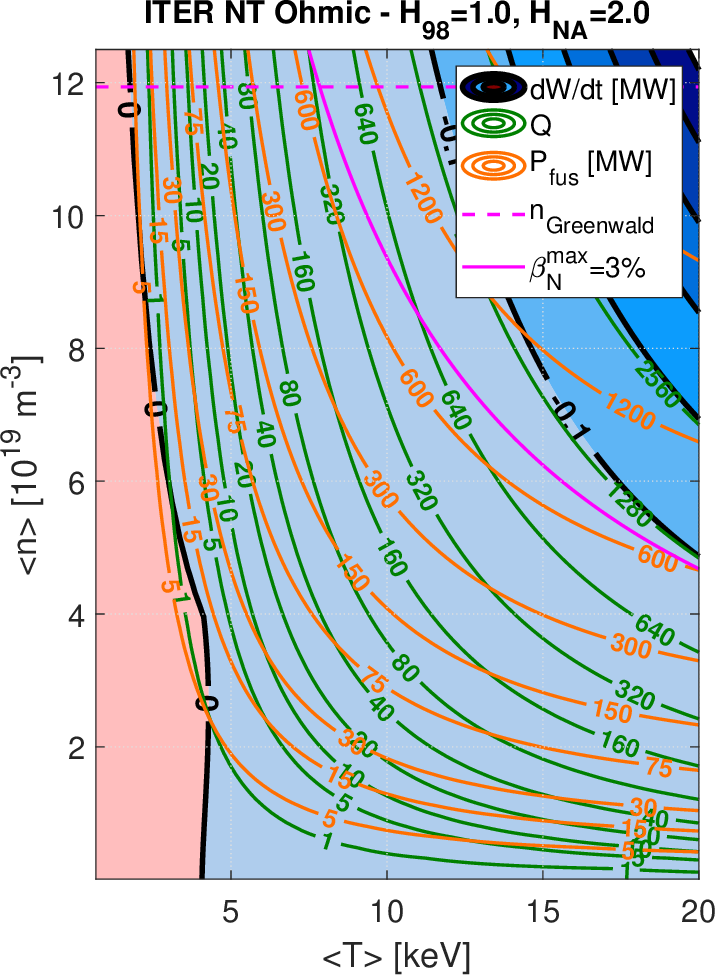}}
    \end{subfigure}
    \begin{subfigure}[]
    {\includegraphics[width=0.24\textwidth]{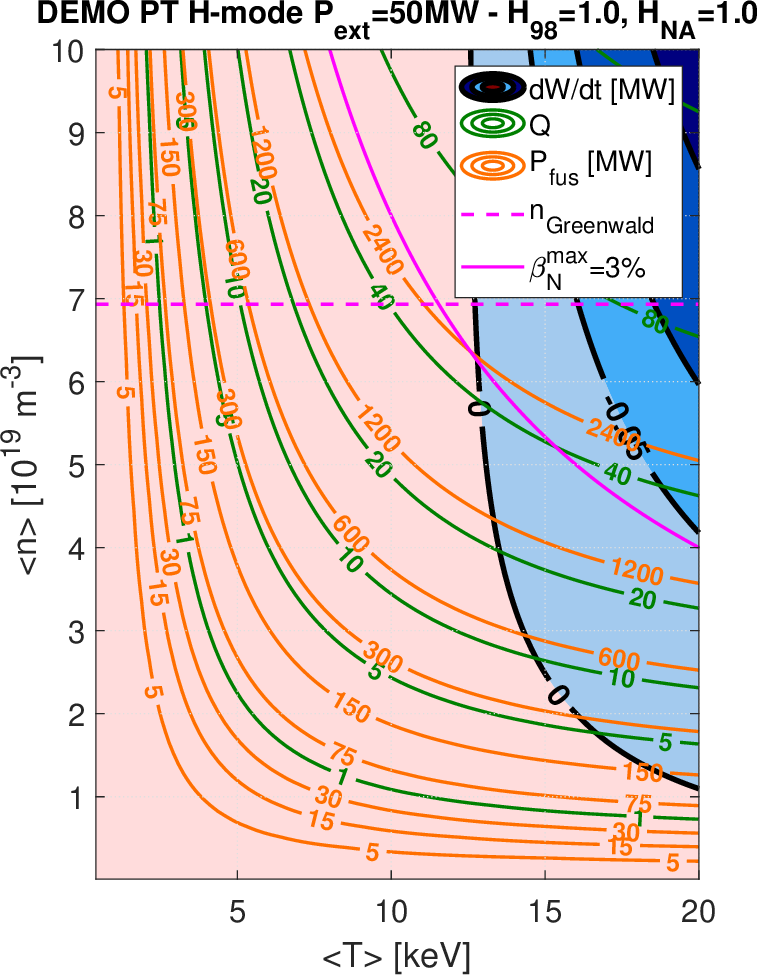}}
    \end{subfigure}
    \begin{subfigure}[]
    {\includegraphics[width=0.23\textwidth]{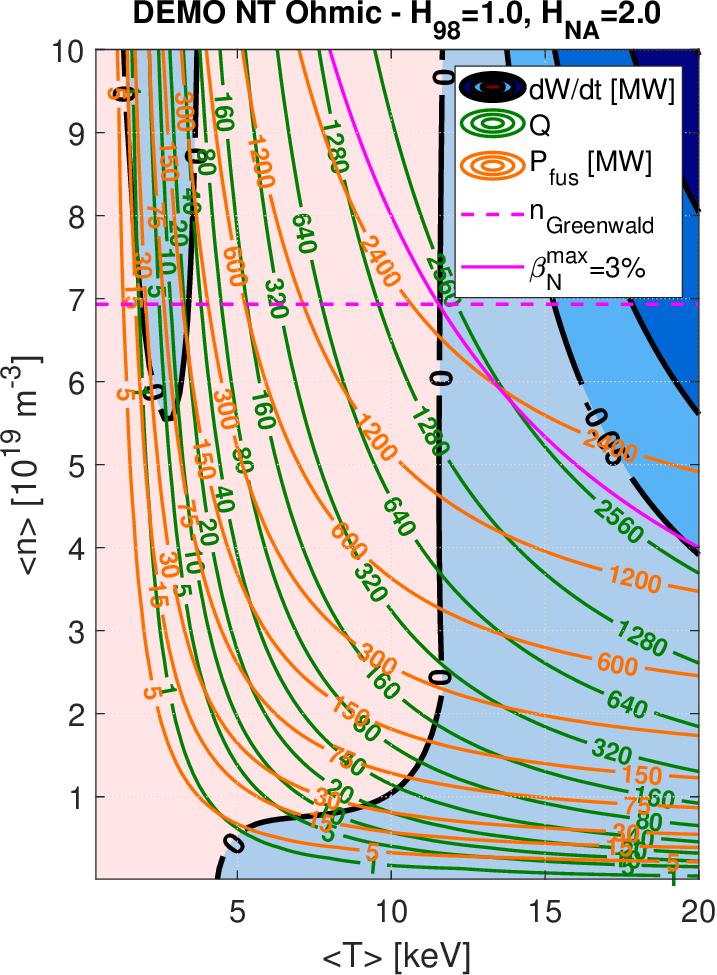}}
    \end{subfigure}
    \caption{Contour plots of $dW/dt$ in temperature-density space for ITER PT (a) and NT (b) scenarios, and for DEMO PT (d) and NT (e) scenarios. The various contour lines have the same meaning as in figure \ref{MANTA}.}
    \label{ITER}
\end{figure*}

We start by comparing NT MANTA scenarios with different energy confinement time scaling laws and enhancement factors. In figure \ref{MANTA} we show operating space plots for scenarios with different levels of external heating. 
In both rows, we take the confinement enhancement factor $H_{NA}=2.0$ for the Ohmic part of the confinement time scaling. These plots show $dW/dt$ over a temperature-density space. The conditions that the device would actually achieve in steady state are represented by the $dW/dt=0$ line. We also display contours of fusion gain $Q$ and total fusion power ($P_{fus}=5P_\alpha$) as well as the Greenwald density \cite{Greenwald_2002} and Troyon $\beta$ limits \cite{Troyon_1988} (i.e. $\beta_N\simeq3\%$) that bound the region where safe operation can be more easily achieved. 

The first row of figure \ref{MANTA} shows that, with H-mode-like confinement, the Ohmic scenario is preferable over the heated cases because the same fusion power can be reached at much larger fusion gain, enabling a plant with higher electrical output. For instance, if we compare figure \ref{MANTA}(a) and \ref{MANTA}(c), we observe that, just below the Greenwald limit, both the cases reach the same fusion power $P_{fus}\simeq1.0\,GW$. However, the Ohmic scenario has a fusion gain of $Q\simeq500$, while the case heated with external power of $P_{ext}=40\;MW$ has $Q\simeq30$. We explain these observations by noticing that, at the chosen operation point, the external heating is negligible compared to the $\alpha$ heating. The only place where $P_{ext}$ cannot be neglected is the denominator of the fusion gain $Q$ (as here it is not summed with $P_{\alpha}$). This is also consistent with equation \eqref{D} as we see that even the Ohmic heating alone can achieve temperatures above the optimal on axis temperature of $6.4\,keV$. To support the validity of 0D analysis, we observe that the results of figures \ref{MANTA}(b) and \ref{MANTA}(c) agree with \cite{Wilson_2024}. In \cite{Wilson_2024}, the authors show that $900\,MW$ of fusion power can be achieved with both $20\,MW$ and $40\,MW$ of external power. Here, we show that may also be possible without external heating. While we argue in appendix B that H-mode-like confinement is the most appropriate for NT plasmas, we can also study the impact of a more pessimistic confinement scaling. To do so, we use the L-mode scaling with an enhancement factor of $H_{89}=1.4$, which  TCV and DIII-D results indicate is very conservative. In the second row of figure \ref{MANTA}, we observe that the Ohmic scenario achieves a similar $P_{fus}$ as the heated case with $P_{ext}=10MW$, but less fusion power than the case with $P_{ext}=40MW$. However, the Ohmic scenario is still able to reach much larger fusion gain $Q\sim40$.

Next we will compare PT H-mode scenarios with NT Ohmically heated ones. We will first consider the PT H-mode scenario envisioned for SPARC (with $P_{ext}=11\,MW$ of ICRF heating) against a theoretical Ohmically heated NT scenario. To produce the NT scenario, we simply flip the triangularity of the PT SPARC scenario (holding the other parameters in table \ref{param} constant). This change in triangularity affects equation \eqref{0D_general} through a change in Ohmic power and in that we use different enhancement factors in the scaling laws for energy confinement time. The results are shown in figure \ref{SPARC}. At SPARC's nominal density ($\langle n\rangle=31\times10^{19}m^{-3}$), the PT H-mode scenario achieves a volume averaged temperature $\langle T\rangle=7.5$ keV, a fusion power $P_{fus}=150\,MW$ and a fusion gain $Q\sim12$. We note that these values agree well with the nominal operation point found in \cite{SPARC}, predicted by much more rigorous calculations. If we assume the H-mode scaling is valid for NT, the same fusion power of PT H-mode can be reached in an Ohmically heated NT plasma and at a much higher fusion gain of $Q=80$. Even if an L-mode scaling is used for NT, the picture does not change much. With an L-mode enhancement factor $H_{89}=1.8$, the fusion power and fusion gain remain the same as for H-mode-like confinement. On the other hand, a more pessimistic factor of $H_{89}=1.4$ yields a lower fusion power of $P_{fus}\simeq75\,MW$, but with the same fusion gain of the PT H-mode case. Lastly, the confinement benefits of less heating also apply to PT L-mode scenarios. For example an Ohmic PT L-mode SPARC scenario appears able to achieve $Q\sim2$ but with much less fusion power (and neutron) production.

Thus far, we have considered machines with large magnetic field ($B_T\geq10\,T$), which is known to be beneficial for Ohmic heating. Let us now consider ITER and DEMO, which are large machines with relatively weaker magnetic field ($B_T\simeq5\,T$). In figure \ref{ITER} we show the reference ITER PT H-mode scenario (which includes $40\,MW$ of external heating) and the reference DEMO PT H-mode scenario (which includes $50\,MW$ of external heating) alongside corresponding artificial NT cases with H-mode-like confinement. We note that our analysis reproduces the expected performance of the nominal ITER and DEMO PT H-mode scenarios well. 

For ITER, we see that the NT scenario is unable to reach the same performance as the PT case. This is because the Ohmic power is too weak and the confinement too poor to attain the optimal H-mode on-axis temperature of $T_0=6.4$ keV. On the other hand, we see that the Ohmic NT scenario is preferable in DEMO (as it reaches a similar fusion power as the PT case with much higher fusion gain). Since DEMO is bigger than ITER, it has a longer energy confinement time, a higher $Q$ and less need for external heating, in agreement with equation \eqref{C}.

\paragraph*{Conclusions}

Since NT tokamaks are not constrained to exceed the L-H heating power threshold, they can improve their energy confinement time by reducing their external heating. While reducing the heating power does not enable more fusion power, it can enable  NT tokamaks to achieve higher fusion gain $Q$ and minimize recirculating power. We observed that NT Ohmic operations are an attractive option for MANTA, SPARC and DEMO, while they are detrimental for ITER. These comparisons along with analytic calculations indicated that an Ohmic NT scenario is most attractive for design points with high fusion gain ($Q \gtrsim 10)$, especially when achieved using a strong magnetic field (as opposed to large size).

\section{Acknowledgments}
This work has been carried out within the framework of the EUROfusion Consortium, via the Euratom Research and Training Programme (Grant Agreement No 101052200 - EUROfusion) and funded by the Swiss State Secretariat for Education, Research and Innovation (SERI). Views and opinions expressed are however those of the author(s) only and do not necessarily reflect those of the European Union, the European Commission, or SERI. Neither the European Union nor the European Commission nor SERI can be held responsible for them. This work was supported in part by the Swiss National Science Foundation.

\section{Appendix A}\label{appA}
In this appendix we give explicit expressions for all the terms on the right side of equation \eqref{0D_general}. All the following formulas give power densities expressed in  units of $keV\,s^{-1}\,m^{-3}$. The $\alpha$ power density is given by \cite{BoschHale}
\begin{equation}
    \langle p_{\alpha}\rangle=\langle \frac{E_\alpha}{4}n^2C_0\zeta^{-5/6}\xi^2e^{-3\zeta^{1/3}\xi}\rangle,
    \label{palpha}
\end{equation}
where $E_\alpha=3500\;keV$ is the energy of the $\alpha$ particles produced by deuterium-tritium fusion reactions, $n$ is expressed in $m^{-3}$, $C_0=6.4341\times10^{-20}$, $\xi=C_1/(T^{1/3})$, $\zeta=1-(C_2T+C_4T^2+C_6T^6)/(1+C_3T+C_5T^2+C_7T^3)$, $C_1=6.661\,keV^{1/3}$, $C_2=1.5136\times10^{-2}\,keV^{-1}$, $C_3=7.5189\times10^{-2}\,keV^{-1}$, $C_4=4.6064\times10^{-3}\,keV^{-2}$, $C_5=1.35\times10^{-2}\,keV^{-2}$, $C_6=-1.0675\times10^{-4}\,keV^{-3}$, $C_7=1.366\times10^{-5}\,keV^{-3}$. The Ohmic power density is
\begin{widetext}
\begin{equation}
    \langle p_{\Omega}\rangle=\langle \eta j^2 \rangle=\frac{\eta_S}{(1-\epsilon^{1/2})^2T_0^{3/2}}\left(\frac{I_p}{a^2\mathcal{F}(\kappa_{95},\delta_{95},\epsilon)}\right)^2\frac{(1+\kappa_0^2)^2}{27 \kappa_0^2}\frac{(1+\alpha_T)^2}{1+0.5\alpha_T},
\end{equation}
\end{widetext}
where $\eta$ is the resistivity, $j$ is the current density, $\eta_S=1.64\times10^8Z_{eff}\,m\,s^{-1}\,A^{-2}\,keV^{2.5}$ is the Spitzer resistivity coefficient, $Z_{eff}$ is the plasma effective charge, $\epsilon=a/R$ is the inverse aspect ratio, $I_p$ is the plasma current in $A$, $\mathcal{F}(\kappa_{95},\delta_{95},\epsilon)=4.1 \times 10^6 (1+1.2(\kappa_{95}-1)+0.56(\kappa_{95}-1)^2)(1+0.09\delta_{95}+0.16\delta_{95}^2)\frac{1+0.45\delta_{95}\epsilon}{1-0.74\epsilon}$ is a function that depends only on geometry and is used to correctly express the current density in terms of $I_p$ \cite{SAUTER2016633} for shaped plasmas, $\kappa_0$ is the elongation on axis, and $\kappa_{95}$ and $\delta_{95}$ are the elongation and triangularity at the flux surface enclosing 95\% of the poloidal flux. The radiated power density is
\begin{equation}
    \langle p_{rad}\rangle=C_BZ_{eff}\frac{n_0^2T_0^{1/2}}{1+2\alpha_n+0.5\alpha_T},
\end{equation}
where $C_B=3.3 \times 10^{-21}\;m^3\,sec^{-1}\,keV^{0.5}$. Here, we are considering only Bremsstrahlung radiation, as it is typically the dominant contribution to the power lost by radiation. The power density lost through transport is extraordinarily difficult to predict from first-principles as it is dominated by turbulence. For this reason, we will use empirical scaling laws based on experimental results for the energy confinement time $\tau_E$ \cite{Goldston_1984}. This enables us to write the loss term as
\begin{equation}
    \langle p_{loss}\rangle=\frac{3n_0T_0}{\tau_E(1+\alpha_n+\alpha_T)},
\end{equation}
where the factor $3$ comes from the assumption that electrons and ions have the same temperature and density. Finally, the external power density $\langle p_{ext}\rangle$ is assumed to be a constant specified value, regardless of the plasma properties.

\section{Appendix B}\label{appB}

\begin{figure}
    \centering
    \begin{subfigure}[]
    {\includegraphics[width=0.48\linewidth]{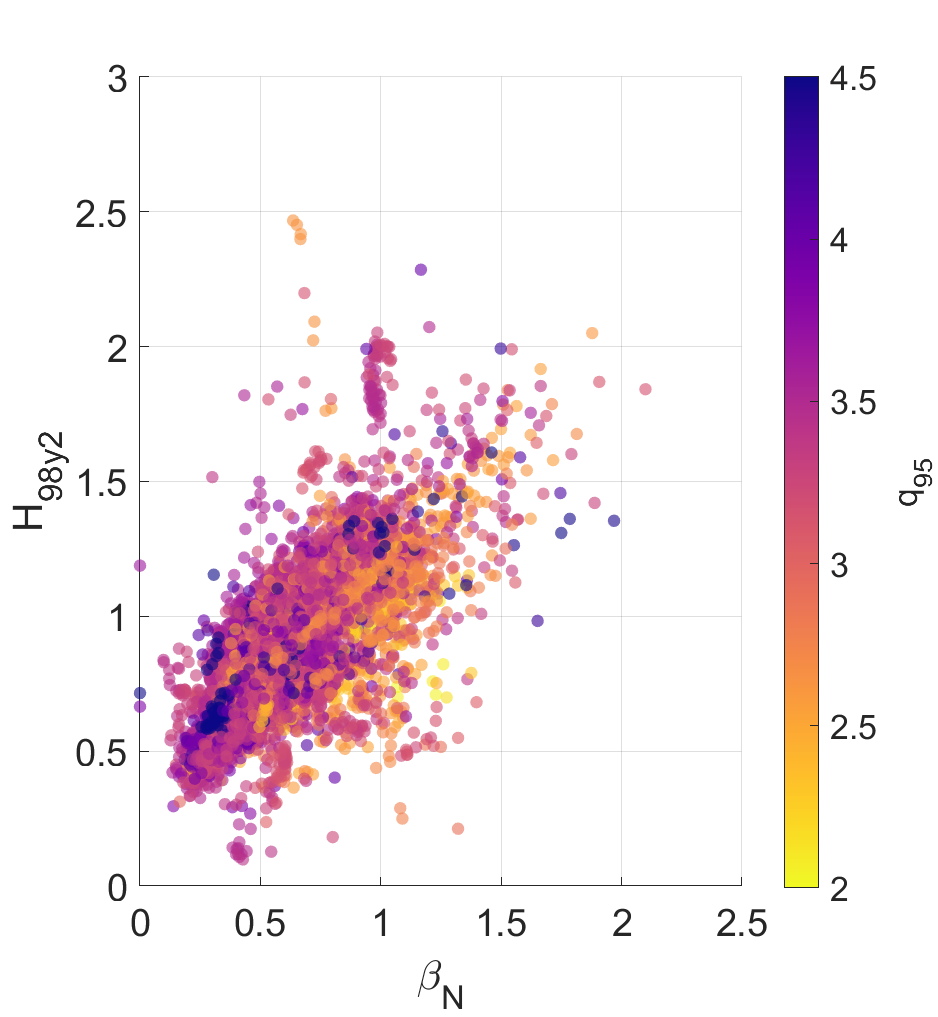}}
    \end{subfigure}
    \begin{subfigure}[]
    {\includegraphics[width=0.48\linewidth]{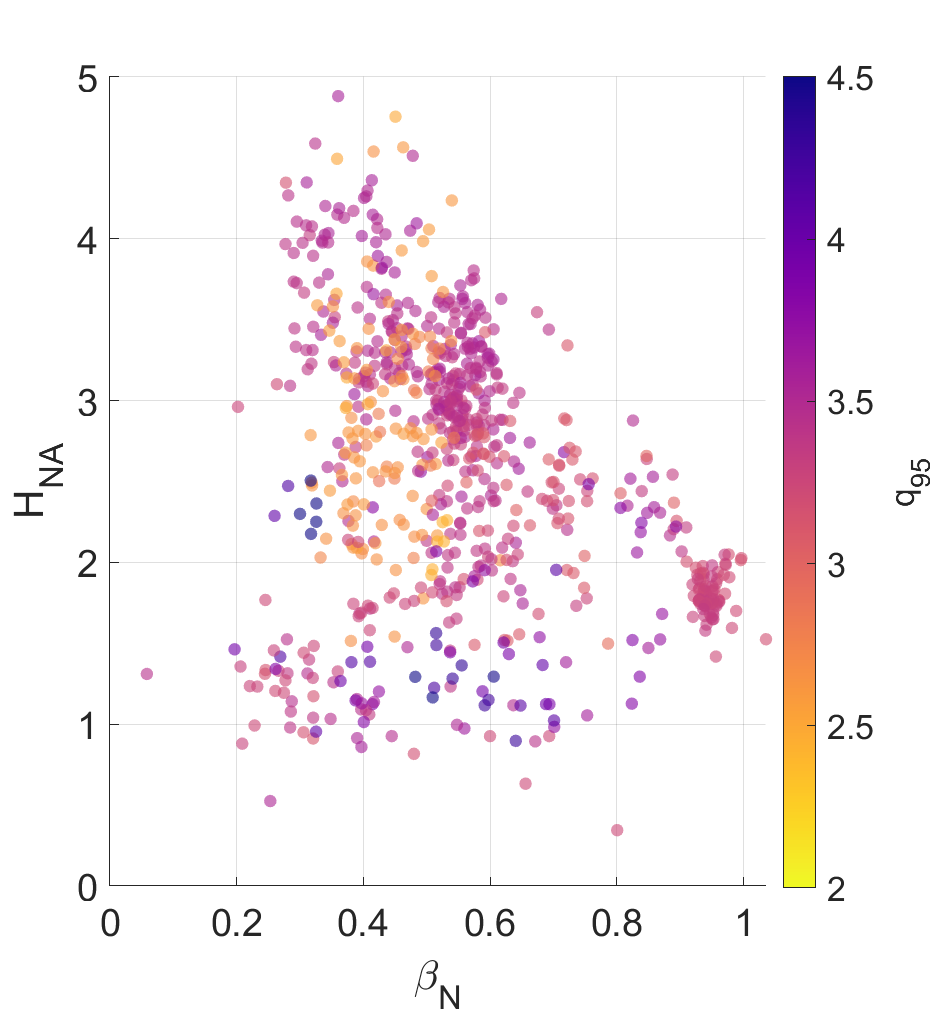}}
    \end{subfigure}
    \caption{(a) ITER98 and (b) Neo-Alcator confinement enhancement factors as functions of normalized $\beta$ and $q_{95}$ from sampled NT TCV shots with external heating (a) and Ohmic heating only (b).}
    \label{histograms}
\end{figure}

For the Ohmic energy confinement time, we use the standard Neo-Alcator scaling \cite{Goldston_1984}
\begin{equation}
    \tau_\Omega=0.007H_{NA}\langle n_{19}\rangle \kappa_{95}aR^2q_{95},
\end{equation}
where $H_{NA}$ is a confinement enhancement factor and $q_{95}$ is the value of the safety factor at the flux surface enclosing 95\% of the poloidal flux. $H_{NA}$ is set to 1 for the best prediction of standard performance, but it can also be used to adjust for considerations not included in the scaling (e.g. NT geometry). For the heated part of the scaling we use either the ITER-98y2 H-mode scaling \cite{ITER98}
    \begin{equation}
        \tau_E^{98}=0.0562H_{98}I_p^{0.93}B^{0.15}n_{19}^{0.41}M^{0.19}R^{1.39}a^{0.58}\kappa_a^{0.78}P^{-0.69}
        \label{tau98}
    \end{equation}
    or the ITER-89P L-mode scaling \cite{Yushmanov_1990}
    \begin{equation}
        \tau_E^{89}=0.048H_{89}I_p^{0.85}B^{0.2}n_{20}^{0.1}M^{0.5}R^{1.2}a^{0.3}\kappa_a^{0.5}P^{-0.5},
        \label{tau89}
    \end{equation}
    where $I_p$ is the plasma current in $MA$, $n_{19}$ and $n_{20}$ are the electron density in $10^{19}m^{-3}$ and $10^{20}m^{-3}$ respectively, $B$ is the magnetic field in $T$, $M$ is the average ion mass in atomic mass units, $R$ and $a$ are the major and minor radii in $m$, $\kappa_a$ is the elongation at the separatrix, and $P=P_\alpha+P_\Omega+P_{ext}$ is the total heating power in $MW$. It is important to note that the negative exponent in the heating power factor means that less external heating improves confinement. Like $H_{NA}$, the confinement enhancement factors $H_{98}$ and $H_{89}$ are scalars representing how much better or worse confinement is with respect to the nominal scaling law (i.e. $H_{98} = 1$ or $H_{89} = 1$).  Since no existing scaling law incorporates NT with sufficiently broad applicability, one is forced to account for NT by choosing representative confinement enhancement factors. These three confinement factors represent the biggest uncertainty in predicting how a NT plasma will behave. A large number of NT discharges produced on TCV (figure \ref{histograms}(a)) and DIII-D \cite{Paz-Soldan_2024,nelson2024_2} indicate that the H-mode scaling of equation \eqref{tau98} with $H_{98}=1$ is a good fit for the experimental performance of heated NT plasmas. However, we will also consider a range of possiblities by using the ITER-89P scaling with different values of $H_{89}$. Regarding the Ohmic confinement time, a preliminary analysis of the TCV NT database suggest that $H_{NA}=2$ can be easily achieved (figure \ref{histograms}(b)) and thus this is our choice for our analysis.

\bibliography{refs}

\end{document}